\documentclass[twocolumn]{aastex631}

\newcommand{\parderiv}[4]{\bigg(\frac{\partial #1}{\partial #2}\bigg)_{#3, #4}}
\newcommand{\deriv}[2]{\frac{\partial #1}{\partial #2}}

\newcommand{\R}{\rm{R}_0^{-1}}

\newcommand{\Mearth}{\rm{M}_{\oplus}}

\newcommand{\Mjup}{\rm{M}_J}

\newcommand{\Yatm}{Y_{\rm atm}}

\newcommand{\Teff}{T_{\rm eff}}
\newcommand{\Zsol}{Z_{\odot}}
\newcommand{\kbbar}{k$_B$ baryon$^{-1}$}
\newcommand{\Mz}{\rm{M}_z}
\newcommand{\Mc}{\rm{M}_c}
\newcommand{\Mj}{M$_J$}

\newcommand{\apple}{\texttt{APPLE}~}

\graphicspath{{./}{figures/}}

\begin{document}

\title{Jupiter Evolutionary Models Incorporating Stably Stratified Regions}

\correspondingauthor{Roberto Tejada Arevalo}
\email{arevalo@princeton.edu}

\author[0000-0001-6708-3427]{Roberto Tejada Arevalo}
\affiliation{Department of Astrophysical Sciences, Princeton University, 4 Ivy Lane,
Princeton, NJ 08544, USA}
\author[0000-0001-6635-5080]{Ankan Sur}
\affiliation{Department of Astrophysical Sciences, Princeton University, 4 Ivy Lane,
Princeton, NJ 08544, USA}
\author[0000-0001-8283-3425]{Yubo Su}
\affiliation{Department of Astrophysical Sciences, Princeton University, 4 Ivy Lane,
Princeton, NJ 08544, USA}
\author[0000-0002-3099-5024]{Adam Burrows}
\affiliation{Department of Astrophysical Sciences, Princeton University, 4 Ivy Lane,
Princeton, NJ 08544, USA}
\affiliation{Institute for Advanced Study, 1 Einstein Dr, Princeton, NJ 08540, USA}

\begin{abstract}
We address the issue of which broad set of initial conditions for the planet Jupiter best matches the current presence of a ``fuzzy core" of heavy elements, while at the same time comporting with measured parameters such as its effective temperature, atmospheric helium abundance, radius, and atmospheric metallicity. Our focus is on the class of fuzzy cores that can survive convective mixing to the present day and on the unique challenges of an inhomogeneous Jupiter with stably-stratified regions now demanded by the \textit{Juno} gravity data. Hence, using the new code \texttt{APPLE}, we attempt to put a non-adiabatic Jupiter into an evolutionary context. This requires not only a mass density model, the major relevant byproduct of the \textit{Juno} data, but a thermal model that is subject to interior heat transport, a realistic atmospheric flux boundary, a helium rain algorithm, and the latest equation of state. The result is a good fit to most major thermal, compositional, and structural constraints that still preserve a fuzzy core and that should inform future more detailed models of the current Jupiter in the context of its evolution from birth.
\end{abstract}


\section{Introduction} \label{sec:intro}

The \textit{Juno} \citep{Bolton2017a} and \textit{Cassini} \citep{Matson2003} missions have significantly transformed our knowledge of the current interiors of Jupiter and Saturn, revealing the presence of high heavy element enrichment at depth and the possible presence of convectively stable regions and extended heavy-element cores \citep{Fuller2014a, Fuller2014b, Wahl2017, Folkner2017, Nettelmann2017, Iess2019, Militzer2019, Debras2019, Durante2020, Mankovich2021, Nettelmann2021, Militzer2022, Militzer2024}. 
In addition, over the past decade exoplanet research is starting to reveal the masses, sizes, and demographic characteristics of thousands of exoplanets \citep[e.g., see][]{Winn2015a, Fulton2017, Petigura2018, Rosenthal2021}. 
Furthermore, missions such as \textit{JWST} \citep{Gardner2006} and future ventures, including \textit{PLATO} \citep{Rauer2014}, and \textit{ARIEL} \citep{Tinetti2021}, are poised to launch a new era of exoplanet exploration.
This ferment suggests that it is timely to attempt a deeper understanding of the structure and evolution of Solar System gas giant planets, in part to advance our knowledge of gas giant exoplanet formation and evolution.

\begin{figure}[ht!]
\centering
\includegraphics[width=0.45\textwidth]{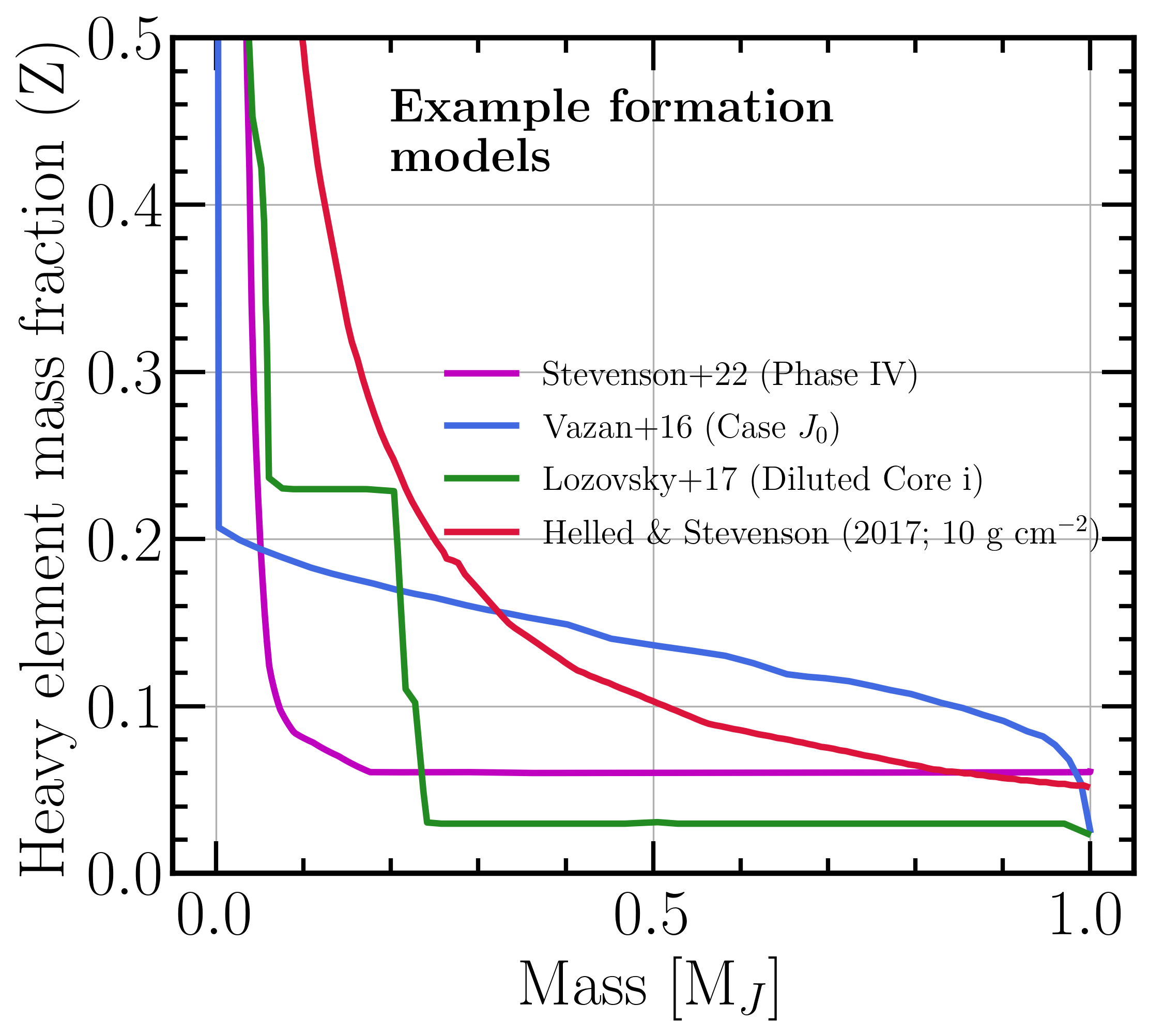}
\includegraphics[width=0.45\textwidth]{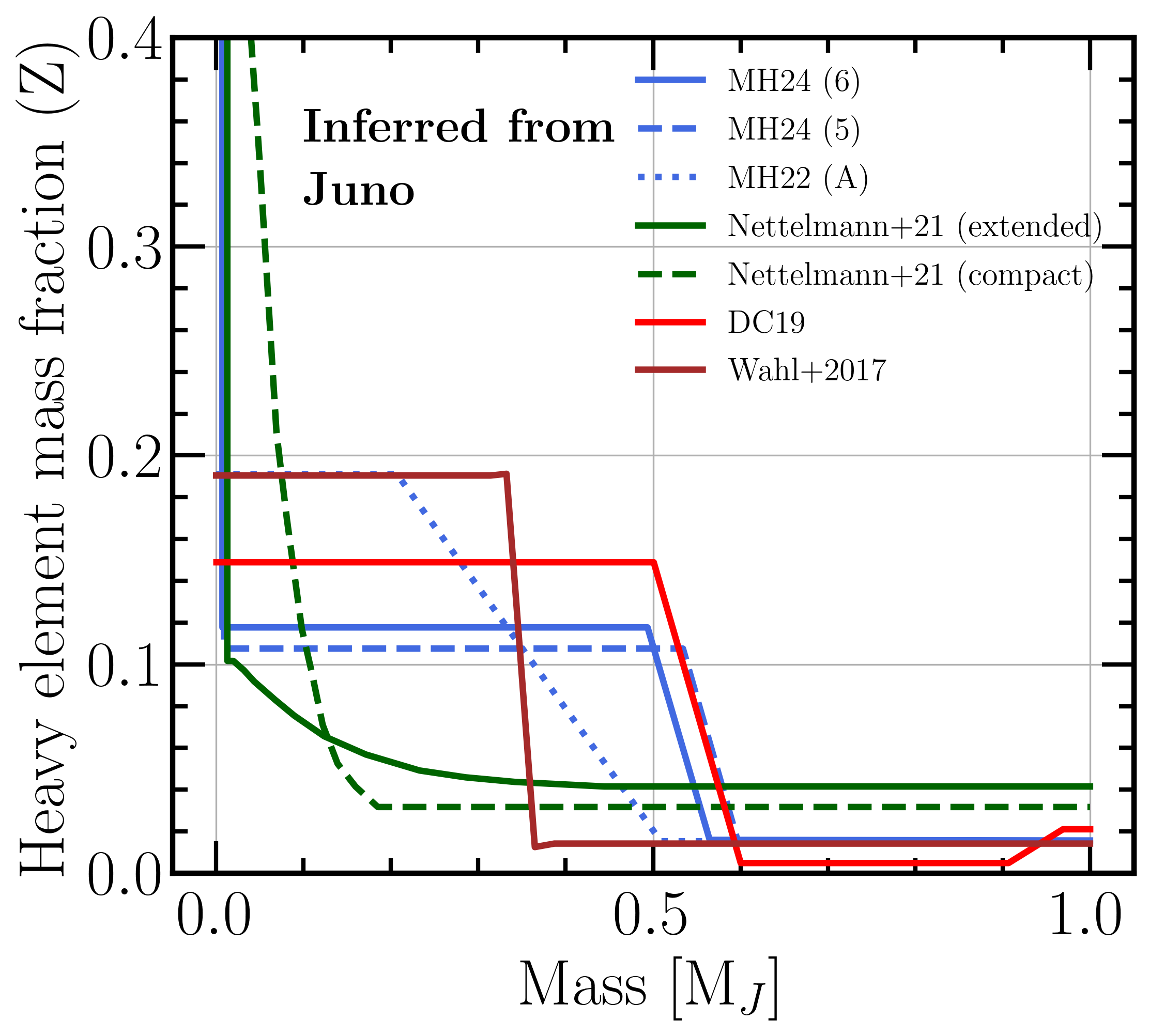}
\caption{Top: Examples of the distribution of $Z$ for formation models from \cite{Stevenson2022}, \cite{Vazan2016}, \cite{Lozovsky2017}, and \cite{helledstevenson2017}. The profile names are also given since these authors explored multiple models. These formation distributions are taken at the time when the proto-Jupiters reached final mass according to the authors. Bottom: Current composition gradients as inferred by the \textit{Juno} gravity data from \citet[][MH24]{Militzer2024}, \citet[][MH22]{Militzer2022}, \cite{Nettelmann2021}, \citet[][DC19]{Debras2019}, and \cite{Wahl2017}, as adapted by \cite{Helled2022b}. The heavy element distributions inferred by various authors using the \textit{Juno} data employ different methods, equations of state, and assumptions concerning the current thermal profile of Jupiter. These differences can potentially account for the differences seen in the inferred $Z$ profiles. Hence, even though \textit{Juno} provided precise values of the $J_{2n}$ gravity moments, there remains a large range of potential $Z$ distributions that might explain the \textit{Juno} data.}
\label{fig:fig1}
\end{figure}

Recent discoveries of extended heavy element ($Z$) regions within Jupiter and Saturn may offer insight into their formation and evolution. 
The top panel of Figure~\ref{fig:fig1} illustrates four examples of initial composition profiles from formation models of Jupiter, as presented by \cite{Vazan2016}, \cite{Lozovsky2017}, \cite{helledstevenson2017}, and \cite{Stevenson2022}.
The bottom panel displays the current inferred heavy-element distribution of Jupiter, as derived from \textit{Juno} gravity data by \cite{Militzer2024}, \cite{Militzer2022}, \cite{Nettelmann2021}, \cite{Debras2019}, and \cite{Wahl2017}. 
These present-day heavy-element distributions of Jupiter encompass a wide range of possibilities that agree with the \textit{Juno} data, though only a limited number of evolutionary models have incorporated formation-based initial conditions \citep{muller2020, Stevenson2022}. 
Evolutionary models informed by formation models have faced challenges in reproducing the heavy-element profiles observed by \textit{Juno} \citep{muller2020}. 
More recently, \cite{Stevenson2022} provided formation models that preserve until the current epoch a dilute core with an extent between 0.33 $R_J$ and 0.44 $R_J$, in rough agreement with the formation models of \cite{muller2020} and \cite{Lozovsky2017}. Particularly, the initial thermal profiles from formation models frequently lead to homogeneous composition profiles. The initial high interior entropies in the formation profiles used by \cite{muller2020} led to extensive convective mixing throughout much of the extended core region over Gyr timescales. 

Thus far, evolutionary studies incorporating Jovian extended cores have left open the questions of 1) which broad set of initial conditions are best suited to match the current presence of a fuzzy core and measured parameters and 2) the character of helium phase separation expected to occur in Jupiter \citep{Stevenson1975, Wilson2010, HubbardMilitzer2016, Mankovich2016, Mankovich2020, Nettelmann2024, Howard2024}. 
We note that past work involving evolutionary models with fuzzy cores \citep[e.g.,][]{Vazan2018, Stevenson2022} did not account for the evolutionary effects of helium rain on Jupiter's atmospheric helium abundance.
To begin to address the question of the possible preservation of a fuzzy core from birth, \citet{Knierim2024} showed that cooler interior entropy ($S$) profiles might maintain an extended initial $Z$ profile over evolutionary timescales.  
In contrast, hotter interior $S$ profiles lead to convective mixing and homogenization over shorter timescales due to the expansion of the outer convective zone. Those authors concluded that the primordial entropy profile is the dominant factor shaping the evolution of the composition profile and broader planetary structure. 
Their conclusions inform formation models and evolutionary calculations, but leave open the question of how Jupiter's current effective temperature and atmospheric helium mass fraction can be explained.

In this paper, we address these two aspects simultaneously. 
We conduct evolutionary models to determine general initial distributions of heavy elements that can best match Jupiter's current effective temperature \citep[$\Teff = 125.57$ K;][]{Li2012}, the atmospheric helium abundance as measured by the \textit{Galileo} entry probe \citep[$\Yatm = 0.234 \pm 0.005$;][]{vonzahn1998}, the outer metallicity abundance of roughly three solar ($\sim$3$\Zsol$) as measured by \textit{Voyager}, \textit{Galileo}, and \textit{Juno} \citep[see][]{HelledHoward2024, Guillot2022, Li2020}, and the current equatorial radius of Jupiter \citep[71,492 km;][]{Seidelmann2007}, while still preserving an extended heavy-element core. 
These values are summarized in Table~\ref{tbl1}. 
This work presents the first evolutionary models of Jupiter with extended cores with helium rain, updated atmospheric boundary conditions that account for ammonia clouds, irradiation, and helium abundance dependence \citep{Chen2023}, the most current hydrogen-helium (H-He) and $Z$ equations of state (EOS) \citep{Chabrier2021, Haldemann2020}, and the best available pressure, temperature, and $Y$-dependent H-He miscibility curves \citep{Lorenzen2009, Lorenzen2011, Schottler2018}. 

In \S\ref{sec:methods}, we describe our evolutionary code, \apple \citep{Sur2024a}, our updated boundary conditions \citep{Chen2023}, the EOSes used for the envelope and compact core, and the H-He miscibility curves used for all models \citep{Tejada2024a}. 
Throughout this work, we use the term ``compact core" to refer to the inner rocky, ``$Z = 1$" region and use ``fuzzy,'' ``extended,'' or ``dilute,'' core interchangeably to indicate the region of the envelope with enhanced $Z$ and/or where $dZ/dm < 0$, where $m$ is the interior mass. 
In \S\ref{experiments}, we provide a set of models with varying initial conditions.  We vary the total entropy profile and the sizes of the extended cores and determine which initial conditions best preserve a fuzzy core, while approximately matching Jupiter's current effective temperature and radius.  
Section \S\ref{sec:results} presents one favored Jupiter model that closely matches the observables in Table~\ref{tbl1} and describes the parameters necessary to obtain this model. We provide concluding remarks in \S\ref{conclusion}. 

\begin{deluxetable}{cc}[ht!]
\tablewidth{0pt}
\tablecaption{Present-day Jupiter Observables Considered}
\tablehead{
\colhead{Parameter} & \colhead{Value}
}
\startdata
$\Teff$ [Kelvin] & $125.57 \pm 0.07$ \\
$\Yatm$ & $0.234 \pm 0.005$ \\
$\Zsol$ (out) & 1.5--5 ($\sim$3) \\
Equatorial Radius [km] & $71,492 \pm 4$ \\
\enddata
\tablecomments{The effective temperature comes from \cite{Li2012}, the atmospheric helium abundance from \cite{vonzahn1998}, the outer metallicity from various measurements of the atmosphere \citep[][]{Li2020}, and the radius measurements come from \cite{Seidelmann2007}. The metal abundances can range between 1.5 and 5 solar but have a concentration of roughly 3 times solar \citep[Figure 2 of][]{HelledHoward2024}}
\end{deluxetable}
\label{tbl1}

\section{Methods} \label{sec:methods}

The evolutionary models presented throughout are calculated using \apple \citep{Sur2024a} and include the effects of solid-body rotation via conservation of angular momentum \citep[see Equation 1 of ][]{Sur2024a}.
A final spin period of 9:55:29.704 hr is assumed \citep{Yu2009} and we neglect differential rotation. 
All models calculate and compare to the equatorial radius of Jupiter. 
Moreover, all models include the precomputed and irradiated atmospheric boundary condition tables of of \cite{Chen2023}.
These boundary conditions are improvements upon the \citet{Fortney2011} boundary conditions in that they include solar insolation in the detailed atmosphere calculation, the effects of ammonia clouds, and naturally obtain the updated albedo of Jupiter of 0.5 \citep{Li2018}. 
We use the H-He EOS of \citet[][CD21]{Chabrier2021} and mix it via the volume addition law  \citep[see][for details]{Tejada2024a, Chabrier2019, Saumon1995} with the latest multi-phase water EOS tables from \citet[][AQUA]{Haldemann2020}, which includes the supercritical and superionic water EOS of \cite{Mazevet2019}.

The most recent theoretical H-He miscibility work is that of \citet[][herein SR18]{Schottler2018}. Both SR18 and the older (and hotter) miscibility curves of \citet[][herein LHR0911]{Lorenzen2009, Lorenzen2011} are thus far the most flexible $Y$-dependent ab initio H-He miscibility curves publicly available. 
We use both of these miscibility curves in this work and adjust, as has been done in past evolutionary work \citep{Nettelmann2015, Pustow2016, Mankovich2016, Mankovich2020, Howard2024}, the temperatures of their miscibility curves at a given pressure and $Y$ fraction when trying to obtain the current helium abundance of Jupiter's atmosphere. We note that recent experimental work on H-He phase separation \citep{Brygoo2021} disagrees with the theoretical demixing temperatures result of LHR0911 and SR18 \citep[see][for a detailed discussion]{Tejada2024a}. The experimental H-He demixing results of \cite{Brygoo2021} find higher immiscibility temperatures by as much as 7,000 K and 3,000 K compared to the theoretical work of SR18 and LHR0911, respectively.

In the absence of any detailed knowledge of helium droplet formation \citep{StevensonSalpeter1977a, Mankovich2016} in regions of expected H-He phase separation ($\sim$1--3 Mbar), \apple uses advection-diffusion methods to transport helium in the miscibility region toward the inner convective regions \citep[see equation 49 in][]{Sur2024a}, while conserving total helium mass to machine precision.
In all our evolutionary calculations, we identify the corresponding helium mass fraction at immiscibility ($Y_{\rm misc}(P,T)$) at the local temperatures and pressures of each Jupiter profile. 
To determine an immiscible region, we compare the adjusted helium mass fraction ($Y^{\prime} = Y/(Y+X)$, where $X$ is the hydrogen mass fraction) with $Y_{\rm misc}(P,T)$.
If $Y_{\rm misc}(P,T) < Y^{\prime}$, then \apple deploys a modified advection term in the diffusion equation to model helium rain \citep[see section 5.4.2 of][]{Sur2024a}.

The regions unstable to convection are determined via the Ledoux criterion,
\begin{equation} \label{ldx_cond}
    \deriv{S}{r} - \sum_i\parderiv{S}{X_i}{\rho}{P}\deriv{X_i}{r} < 0\, ,
\end{equation}
where the $X_i$ are the helium ($Y$) and heavy-element ($Z$) mass fractions.
A full derivation of this criterion can be found in \cite{Tejada2024a}. The EOS tables and a full description of these derivatives are also found in \cite{Tejada2024a}. 

In convective regions, a mixing-length flux procedure for both energy and composition, as described in \citet{Sur2024a}, is employed. Equation~\ref{ldx_cond} is tested at every mass zone and if the convective condition is satisfied, we add the convective flux to the conductive heat flux. However, as discussed in \cite{Gabriel2014}, determining the stable-convective boundary is not a trivial problem and the determination and effects of the stable-convective boundary have been the subject of intense study in stellar evolution circles. Our application of the Ledoux condition assumes that the convective boundary zone is located between the last convective zone and the first stable zone, introducing a discontinuity in the Ledoux condition derivatives. The challenges in determining the stable-convective boundary regions are many and there is currently no best practice adopted by the stellar or planetary evolutionary community. This is an important challenge that the stellar and planetary evolution communities should resolve in the near future.

In stably-stratified regions, the conductive thermal and compositional fluxes are calculated using the \citet{french2012} diffusivities,
as modified by \citet{Sur2024a}. 
Given its current uncertain nature and likely transient effects, we do not include semi-convection in any models in this work. The effects of semi-convection could be minimal in Jupiter over evolutionary timescales, as observed by \cite{muller2020}, but this topic certainly warrants future scrutiny. Each timestep is determined to ensure the temperature, helium fraction, and heavy-element fraction do not change in a zone by more than 1\% and 750 spatial zones are employed.
All models in this work conserve cumulative energy to within $1\%$, using Equations 58 and 59 in \cite{Sur2024a}.
We conducted resolution tests with mass zones spanning 250 and 1000 zones and temporal resolution between 0.5\% and 5\%. We find that models with more mass zones and temporal resolution require significantly more computing time than we suggest is warranted by the corresponding modest improvement in energy conservation ($\sim0.2\%$).

\section{Preservation of Initial Heavy-Element Extended Cores} \label{experiments}

This section explores the general contexts in which an extended ``fuzzy" Jovian core of heavy elements might survive until the present day. 
As illustrated in the bottom panel of Figure~\ref{fig:fig1}, the range of fuzzy core structures inferred from the \textit{Juno} data is quite broad. 
\citet{Debras2019}, \citet[][MH22]{Militzer2022}, and \citet[][MH24]{Militzer2024}, infer flat $Z$ profiles out to a steep $Z$ gradient near $\sim$0.6 R$_J$, where R$_J$ is the radius of Jupiter.
If composition profile gradients are overrun by convection, they will then be mixed and will not survive over Gyr timescales.  
A narrow $Z$ gradient, flanked by flat interior and exterior profiles, presents a physical challenge, since such a configuration would naturally arise only under conditions of homogenizing convection.
A fully convective, adiabatic initial envelope, which would homogenize the planetary structure, is inconsistent with the presence of a fuzzy core. 
Therefore, we initialize the envelope as stably stratified, imposing an initial positive entropy gradient to achieve this. 
Hence, we investigate the general conditions under which convection can be suppressed, allowing a fuzzy core and its associated stabilizing $Z$ gradients to persist.

In \S\ref{flat_test} we explore how a class of initial steep $Z$ profile gradients in the middle of the planet surrounded by flat $Z$ distributions might survive Gyr timescales of thermal, structural, and compositional evolution. 
In \S\ref{flat_test_2}, we explore the dependence of the $Z$ profile gradient on internal entropy profiles by varying from cold to hot internal entropies.
The initial models in \S \ref{flat_test} and \S \ref{flat_test_2} are constructed with linear profiles along the mass coordinates of the structure. For each model, there are two flat linear profiles surrounding a steep linear heavy element profile. We maintain three linear, piecewise profiles for the entropy, but introduce slopes in the regions of flat $Z$ profiles to sustain a fully stable initial condition.
We present these models graphically in Figures~\ref{fig:fig2}--\ref{fig:fig5}. 
This is followed in \S\ref{gradual_test} by an investigation into how fuzzy cores with a different class of initial $Z$ profiles which gradually decrease in $Z$ abundance from the inside out might fare over time. 
This different class of initial conditions resembles the more smoothly continuous present-day Jovian $Z$ distributions inferred by \citet{Nettelmann2021}.
We emphasize that we are not directly addressing in this study planet formation scenarios that might produce such profiles. 

All the models explored in \S\ref{flat_test}--\S\ref{gradual_test} are given a total heavy-element mass of 27 $\Mearth$. 
The choice of 27 $\Mearth$ for these ``flat-top'' experiments are motivated by the recent suggestions of \citet{Militzer2024}, who found that a total heavy element mass of 27 $\Mearth$ is consistent with their flat-top fits to the Juno data. 
Nevertheless, any heavy element total mass would demonstrate the effects shown in \S 3.1 and \S 3.2..
Of this total, 2.5 $\Mearth$ are put in a compact core made up of $50\%$ water and $33.33\%$ post-perovskite (MgSiO$_3$), with the remainder iron using the EOS used in \cite{Zhang2022}, which is Keane's EOS \citep{Keane1954, Stacey2004} expanded to high pressures and temperatures for this work. 
For simplicity, water is the only heavy species in the envelope. 
Furthermore, all initial models assume an initial constant helium abundance relative to hydrogen at the protosolar value of $Y^{\prime} = Y/(X + Y) = 0.277$, as calculated by \cite{Bahcall2006}.

\subsection{The Survival and Consequences of Flat-Top Fuzzy Cores}
\label{flat_test}

\begin{figure*}[ht!]
\centering
\includegraphics[width=\textwidth]{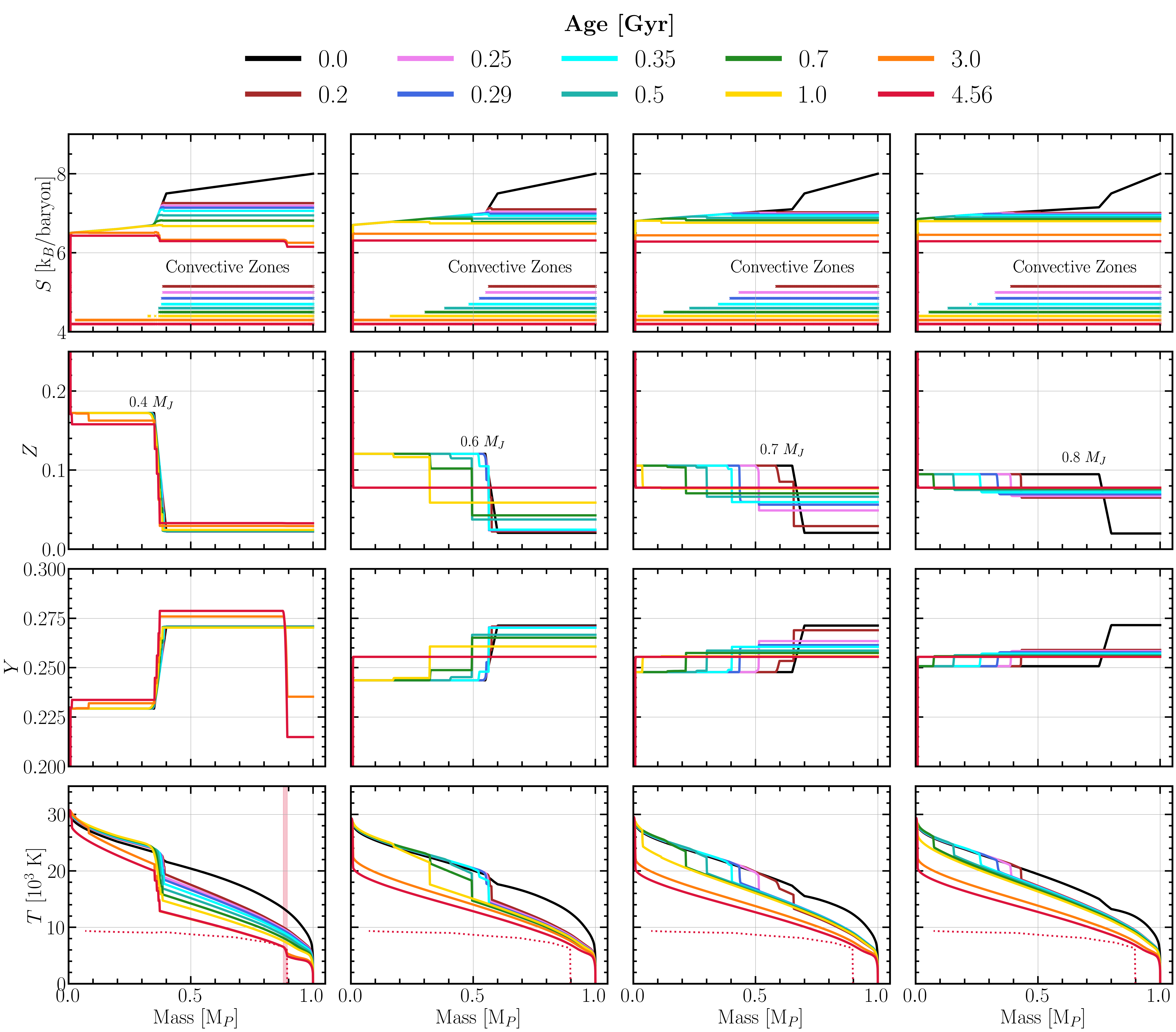}
\caption{Example evolutionary calculations of four mock models with varying initial sizes of the extended core. From left to right, the models increase in extent from 0.4 to 0.8 \Mj. All models have a total envelope heavy-element mass of 27 $\Mearth$ with 2.5 $\Mearth$ in the compact core ($Z = 1$). The four models in this study assume the same initial entropy of 7.5 \kbbar at the outer edge of the fuzzy core but shift in location to match the location of each stable region. The surrounding flat $Z$ profile regions have been initialized with stabilizing $S$ profile gradients. None of these models should be taken as the preferred models of Jupiter. Rather, these models are meant to explore the contexts for the survival of the given fuzzy cores to late times. Since these are example models only, we have used the original LHR0911 miscibility curves to demonstrate helium rain. The first row depicts the entropy profile and the inward extension of the convective zone as each structure cools. At later ages, when the outer region has cooled more than the inner stable region, the convective zones surround the steep stable regions, leaving behind so-called ``staircases" \citep[see also][]{Vazan2018}. The second row shows the evolution of the $Z$ profile. Generally, these figures show that shallower (i.e., located closer to the surface) extended cores experience more convective mixing than deeper extended cores. The third row shows the evolution of the helium abundance profile. Due to the conservation of mass and the setting of the hydrogen/helium mass ratio equal to the proto-solar value \citep[$Y/(X+Y) = 0.277$;][]{Bahcall2006}, the helium abundance increases with decreasing heavy metal abundance. The last row shows the evolution of the temperature profile with the helium rain region of the 0.4 $\Mjup$ model shaded in red. Note that all but the 0.4 $\Mjup$  model mixes entirely by 4.56 Gyrs. As a result, the temperatures in the convective envelope of the 0.4 $\Mjup$  model are colder than the rest at this age. At early ages, however, the temperature profiles of this model are higher than the rest, since it can access more internal heat due to a deeper stable region. Deeper extended cores allow more heat to be available to the outer convective zone and this maintains higher outer envelope temperatures.}
\label{fig:fig2}
\end{figure*}

\begin{figure}[ht!]
\centering
\includegraphics[width=0.47\textwidth]{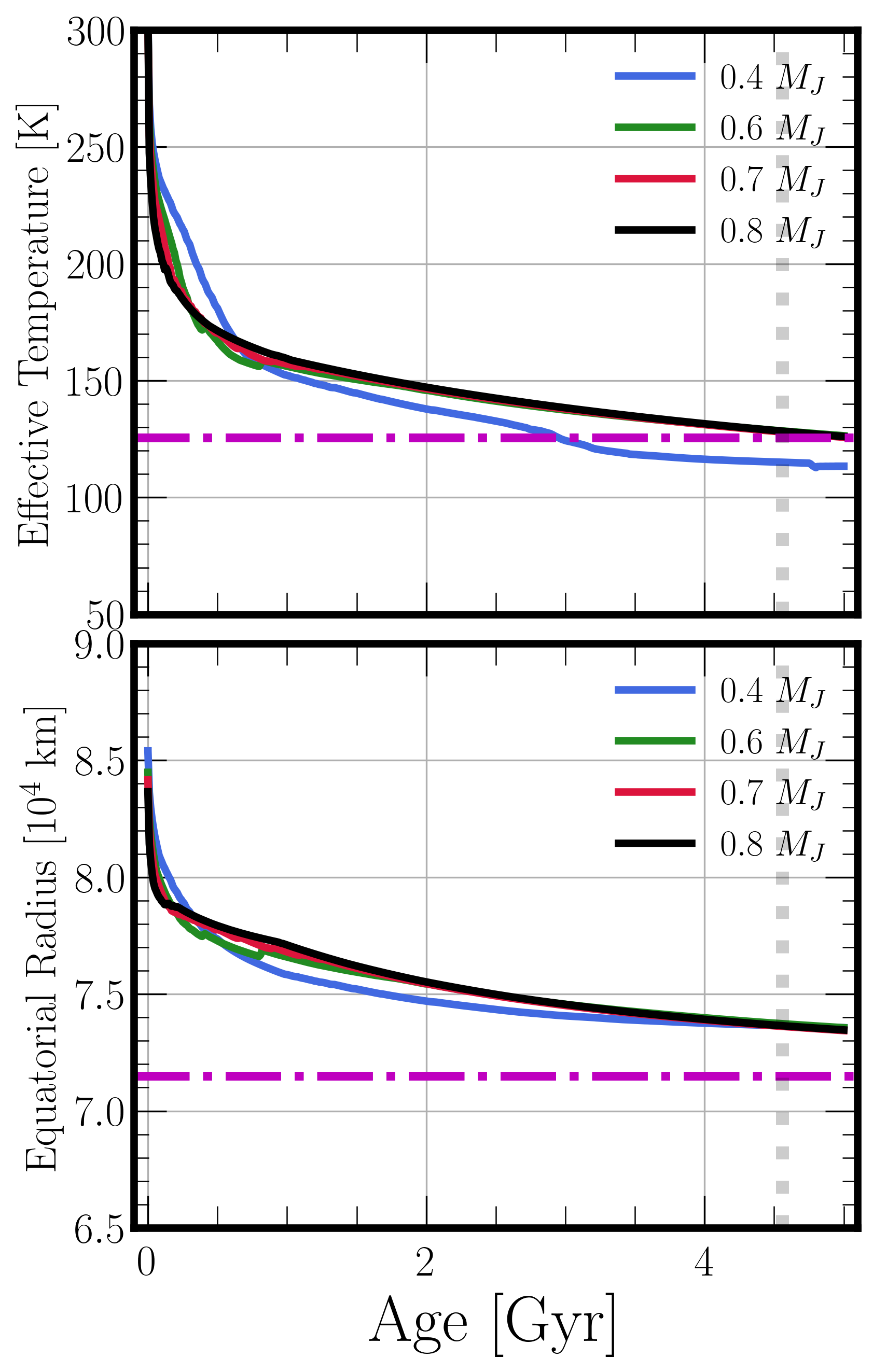}
\caption{Effective temperature (top panel) and equatorial radius (bottom panel) evolution of the models presented in Figure~\ref{fig:fig2}. The models are distinguished by the initial extent of the stable regions corresponding to the columns in Figure~\ref{fig:fig2}. The deepest fuzzy core model, 0.4 \Mj, accesses more heat than other models at early times. At later times, however, this model still cannot efficiently access interior heat, while the other models have convectively mixed and, thus, their interior heats are made available to their exteriors. The effective temperature of this model, therefore, is higher than those of other models at early times, but lower at later times. The violet horizontal line in the top panel marks the observed effective temperature of Jupiter of 125.57 K \citep{Li2012} and the equatorial radius in the bottom panel, as measured by \citep[72,492 km;][]{Seidelmann2007}. Due to the high interior temperatures, the radii remain larger than the equatorial radius of Jupiter in these mock evolutionary models. The purpose of Figure~\ref{fig:fig2} and this figure is not to obtain the correct radius (for that, see \S{\ref{sec:results}} and discussions thereafter), but rather to illustrate that only extended stable regions located at $\lesssim$ 0.4 $\Mjup$ can maintain a semblance of a stable region.}
\label{fig:fig3}
\end{figure}

The initial heavy metal ($Z$) and entropy ($S$) profiles presented in this section follow the same general trends recently found by \cite{Knierim2024}.
Namely, that initial $Z$ profiles located at larger radii convectively mix\footnote{Our working definition of ``mixing'' are regions where $\partial Z/\partial m  = 0$.}more rapidly than those located deeper in the structure.
From left to right, Figure~\ref{fig:fig2} depicts what happens when the initial location of the steep gradient is moved progressively outward.
The convective zones move inward, and these zones convectively mix the extended cores, as shown in the second row of Figure~\ref{fig:fig2}.
Models that mix throughout access all the heat of the envelope, leading to higher effective temperatures.
In Figure~\ref{fig:fig3}, we show that the models with larger extended cores experience lower effective temperatures due to inefficient heat transport from their interiors at early times. 
The model with its initial fuzzy core extending out to 0.4 $\Mjup$ (shown in blue) can hold its extended core over Gyr timescales and shows higher effective temperatures at early times since it can access more of this interior heat.
The evolution of the equatorial radius is shown in the bottom panel of Figure~\ref{fig:fig3}. 
These example models illustrate the dependence of the effective temperature on the initial $Z$ and $S$ profiles.
The 0.4 $\Mjup$ model in Figure~\ref{fig:fig3} reaches the current Jovian effective temperature at $\sim$3 Gyrs, and its extended core is preserved only by narrow stable regions enclosed by convective regions with no extended stable zone.
The convective regions can penetrate the $Z$ profile gradient when the entropy profile of the outer region cools to lower values close to those of the innermost stable region of the model.
Our evolution profiles show so-called ``staircases," also observed by \cite{Vazan2018} and \cite{muller2020}. Stellar evolution codes, such as MESA \citep{Paxton2011, Paxton2013} also exhibit similar staircase behavior \citep[see Figure 2 of][]{Paxton2018}. They are not expected to occur in Nature, are not observed in 3-D hydrodynamic simulations, and are not the same as those expected by semi-convection or double-diffusive convection \citep{Leconte2012, Wood2013, Moll2017, Tulekeyev2024}.
The models in this section are not representative of Jupiter's evolution, and for demonstration purposes only, we have used the unmodified LHR0911 miscibility curve, shown in the bottom row of Figure~\ref{fig:fig2}.

\subsection{The Consequences of Initial Entropy}
\label{flat_test_2}

Initial fuzzy cores with higher (or ``hotter") $S$ profiles mix more rapidly due to convection than those with lower (or ``colder") $S$ profiles \citep{muller2020, Knierim2024}. 
From left to right, Figure~\ref{fig:fig4} illustrates this dependence of the $Z$ profile on the $S$ profile. 
The location of the $Z$ profile is maintained in Figure~\ref{fig:fig4} at 0.6 $\Mjup$ to illustrate the $S$ profile effects on the $Z$ profile. To guide the eye, a characteristic entropy at the position of the $Z$ gradient (ranging from 6.5 to 8.5 \kbbar) is marked by black dash-dot lines in the first row.  
The inward advancement of the convective front is inhibited by the $Z$ gradient in colder $S$ profiles, causing the outer convective regions to cool more rapidly without heat replenishment from the interior. 
This outer adiabatic cooling results in lower effective temperatures and smaller radii, as shown in Figure~\ref{fig:fig5}. 
As outer temperatures drop, they intersect the shifted SR18 miscibility curve, as shown in the left-most column. 
As such, preserving these extended cores results in effective temperatures much colder than the present-day Jovian effective temperature. 
As in the previous section, and since these are only example models, we arbitrarily increased the demixing temperatures of the SR18 miscibility curve by +550 K for these example calculations.

\subsection{The Survival and Consequences of Gradually Sloping Fuzzy Cores}
\label{gradual_test}

The shortcomings and challenges of the profiles explored in \S\ref{flat_test} and \S\ref{flat_test_2} motivate the exploration of a different kind of initial distribution. 
The reader should keep in mind that although formation models have provided initial conditions to be used in evolutionary models, these have not in the past been shown to preserve fuzzy cores larger than $\gtrsim$ 0.2 $\Mjup$ \citep{muller2020}. 
Based on our example evolution models described in \S\ref{flat_test} and \S\ref{flat_test_2}, initially steep $Z$ profile gradients extending beyond 0.4 $\Mjup$, as advocated in the models of MH24, do not seem at the same time to easily explain the Jovian effective temperature in an evolutionary context. 
Those initial $Z$ distributions which have an initially stable interface at interior masses greater than 0.5 $\Mjup$ mix too rapidly to provide a stable $Z$ region at the present age, as shown in Figure~\ref{fig:fig2}. 
If such a profile is initialized with colder interior entropy profiles, then the stable $Z$ region will endure (Figure~\ref{fig:fig4}). 
The outer regions, though, will cool rapidly and fail to account for Jupiter’s current effective temperature, as illustrated in Figure~\ref{fig:fig5}. 
These colder models can be provided with higher exterior entropy profiles to achieve higher initial effective temperatures. 
Although such models experience higher initial effective temperatures, they nonetheless cool to effective temperatures below 120 K at later ages. 
One may believe that this problem might be alleviated with higher initial internal temperatures near the steep $Z$ gradient, but the associated elevated internal entropies will mix the profile gradients, as demonstrated in Figure~\ref{fig:fig4}. 
A negative $Z$ gradient can be introduced interior to the steep portion of the profile to keep the extended core from fully mixing.
This maintains the extended core out to 0.5 M$_J$, but the effective temperatures remain below 120 K at 4.56 Gyrs.
The requirement to maintain the steep $Z$ gradient at $M \sim$0.5--0.6 M$_J$ conflicts with the need to sustain high enough temperatures in the convective region to achieve an adequate $\Teff$ for this kind of profile. 
Thus, while a steep $Z$ profile gradient can persist over Gyr timescales under certain conditions, we disfavor such initial configurations due to the requirement to match Jupiter's effective temperature. 

Facing these challenges, we explored a more gradually sloping class of initial conditions. The exercise in \S\ref{flat_test_2} is repeated in Figures~\ref{fig:fig6} and \ref{fig:fig7}, where the initial fuzzy cores are instead gradual throughout the inner envelope. The unmodified miscibility curves of LHR0911 have been used to illustrate helium rain in these example models. 
As in Figure~\ref{fig:fig4}, Figure~\ref{fig:fig6} depicts models for which a characteristic entropy (shown in the first row) ranges from 6.5 to 8.5 \kbbar, demonstrating the effect of the gradual increase of the $S$ profile from left to right. 
The rightmost model described by the highest entropy mixes completely by 0.4 Gyr, while the others maintain their extended fuzzy cores to varying degrees. 
In Figure~\ref{fig:fig7}, we show that the gradual core evolutionary model characterized by 7.5 \kbbar (third column), which indeed preserves a stable fuzzy core, can reach an effective temperature closer to that of Jupiter. 
This class of initial conditions allows for a more gradual evolution of the effective temperature (see Figure~\ref{fig:fig7}). 
As shown in Figure~\ref{fig:fig6}, internal entropies of between 7 and 7.5 \kbbar appear adequate for this kind of initial condition.

Initial internal entropies of 7.5 kb/baryon are in contrast to past ``hot-start" gas giant models. Recently, \cite{muller2020} explored hot ($\sim$9 \kbbar) interior entropies along with cold ($\sim$7 \kbbar) formation models of Jupiter and evolved them to the present epoch. 
The ``Cold extended'' model of \cite{muller2020} and the cold models of \cite{Cumming2018} have entropy profiles consistent with our own. 
The cold profiles of \cite{Cumming2018} and \cite{muller2020} both have $\sim$7 \kbbar \ in their deep stable regions, as do ours. 
We note that depending on the initial entropy profile, the outer entropy can gradually rise to entropies reminiscent of the older ``hot-start" models, close to 9-12 kb/baryon. 
However, these rapidly ($<$ 10 Myr) cool to entropies below 8 kb/baryon. Such an initial entropy distribution can be considered a hybrid of cold and hot start models.

\section{Evolution of Jupiter}\label{sec:results}

To find a favored model with the class of initial conditions discussed in \S\ref{gradual_test}, we conducted a parameter search by varying the total mass of heavy elements, $\Mz$, from 25 $\Mearth$ to 50 $\Mearth$.
For each $\Mz$, the compact core mass ($\Mc$) was varied from 2 to 15 $\Mearth$. 
The ranges of total heavy element masses and compact core masses are motivated by the lower and upper mass limits allowed by the Juno gravity data \citep[see Figure 3 of ][]{Wahl2017}.
For each of these models, we varied the miscibility temperature modifications ($\Delta T$) from -500 K to 500 K  (for LHR0911) and from 0 K to 2,500 K (for SR18). 
Figures~\ref{fig:fig2}--\ref{fig:fig7} show that internal characteristic entropy values of 7.5 \kbbar\ strike a balance between preserving an extended core and yielding appropriate effective temperatures for Jupiter at the present age. 
Thus, each model was initialized with a characteristic interior entropy of 7.5 \kbbar and an outside entropy of 8.2 \kbbar. 
Although a wide variety of initial profile distributions could be constructed and brought close to matching observed data, this particular class of initial $Z$ distributions is naturally suited to match the observables presented in Table~\ref{tbl1}. 
Specifically, it preserves an inner stable region better and provides enough erosion of the extended core via convective mixing to obtain a metallicity of 3$\Zsol$ in the outer regions.

Our preferred model for Jupiter's evolution is presented in Figures~\ref{fig:fig8} and \ref{fig:fig9}. The observed parameters of this model are compared to the data for Jupiter in Table~\ref{tbl2}.
The relative errors of this model's effective temperature and radius are within 0.05\% and 0.2\% respectively, all while $\Yatm$ reaches the value of 0.234.
In this model, helium rain begins at 4 Gyrs. 
Importantly, the miscibility temperatures required to achieve $\Yatm = 0.234$ are +300 K higher than the original LHR0911 immiscibility temperatures. 
At the same time, a significant temperature modification of +2,200 K to the SR18 miscibility curve is required. 
The required miscibility curves of LHR0911 and SR18 are compared to the original model in Figure~\ref{fig:fig10}, with experimental H-He phase separation data from \cite{Brygoo2021} interpolated and plotted for comparison.
Our prediction of a higher miscibility temperature falls within $3\sigma$ of the \cite{Brygoo2021} experimental results.
The elevated immiscibility temperatures are in part a result of the higher profile temperatures in the convective regions due to the presence of heavy metals in the envelope. The difference seen in Figure \ref{fig:fig10} between the shifted LHR0911 and SR18 miscibility curves is a consequence of their different shapes in the $\sim$1$-$4 Mbar pressure region of relevance to helium rain in Jupiter. 
A recent evolutionary calculation that incorporates only H-He in the interior of Jupiter's envelope predicts colder demixing temperatures, by as much as -1,250 K concerning the original LHR0911 miscibility curve \citep[][]{Howard2024}. 
The best-fit Jupiter model of \citet{Mankovich2020} seems to have a homogenous envelope metallicity of only $\sim$one solar, with the correspondingly lower envelope temperatures and lower required shift to the SR18 miscibility curve.  
We conclude that the inclusion of heavy metals in the envelope results in higher temperatures and, therefore, warrants higher H-He demixing temperatures.

Our preferred Jupiter model is the first to incorporate the evolution of initially stable extended cores to match the effective temperature, radius, outer metal abundance, and atmospheric helium abundance with helium rain within reasonable uncertainties. 
One of the subsidiary conclusions we draw from the explorations in \S\ref{flat_test} and \S\ref{flat_test_2} is that steep $Z$ profile gradients surrounded by flat $Z$ profiles left over from formation will lead to effective temperatures that are too low for the present Jupiter. 
This is due to the lack of heat resupply from the interior, since heat transport in these stable regions relies on conduction, which is significantly less efficient than convection. As a result, stable regions located between 0.4 M$_J$ and 0.6 M$_J$ severely inhibit heat transport.

Since the aim of this study was to investigate which initial distributions 1) can retain extended stable cores and 2) can account for Jupiter's effective temperature, atmospheric helium abundance, and equatorial radius, we chose initial distributions better suited for that purpose. 
Past evolutionary models informed by formation models have thus far been unable to reproduce the inferred \textit{Juno} $Z$ profiles \citep{muller2020, Stevenson2022}. The evolutionary models of \cite{Vazan2018} end with an extended core region out to 0.38 M$_J$ but did not attempt to impose the constraint of the measured atmospheric helium abundance, used the older SCvH EOS \citep{Saumon1995}, and did not employ an up-to-date atmospheric boundary. The stable region predicted by the formation and evolutionary models of \cite{muller2020} extends to only 0.2 M$_J$. More recently, \cite{Howard2023b} investigated the H-He EOS dependence of static Jupiter models, also favoring an extended core out to 0.2 M$_J$.  The evolutionary models presented in our work begin with a completely stable structure but also end with an extended core region out to $\sim$0.3 M$_J$.

Our work is distinguished from the works of \cite{Vazan2018}, \cite{muller2020}, \citet{Mankovich2020}, \citet{Howard2024} in the following ways:

\begin{enumerate}
    \item The use of updated boundary conditions which account for irradiation, the Bond albedo, and ammonia clouds \citep{Chen2023}. These boundary conditions are Y-dependent, adjusting to the outer $Y$ abundance during helium rain.
    \item The use of an updated H-He EOS \citep{Chabrier2021} and the incorporation of a multi-phase water EOS \citep{Haldemann2020}.
    \item An exploration of different classes of initial condition profiles for $S$ and $Z$. The various initial conditions explored by \cite{muller2020} were formation models, whereas here we have explored the evolutionary consequences of the flat-top profiles and smooth profiles in the thermal evolution of Jupiter.
    \item The application of the two most recent H-He miscibility curves (LHR0911, SR18) used in the evolution of $Y$ in the presence of $Z$ profile gradients. These H-He miscibility curves are allowed to adapt and adjust to the local pressures, temperatures, and helium abundance.
    \item The deployment of a new evolutionary code \citep[\apple;][]{Sur2024a} with updated microphysics, such as conductivities, helium rain modeling methods, and the inclusion of centrifugal effects on the structures.
\end{enumerate}

\section{Conclusions}
\label{conclusion}

\begin{figure*}[ht!]
\centering
\includegraphics[width=\textwidth]{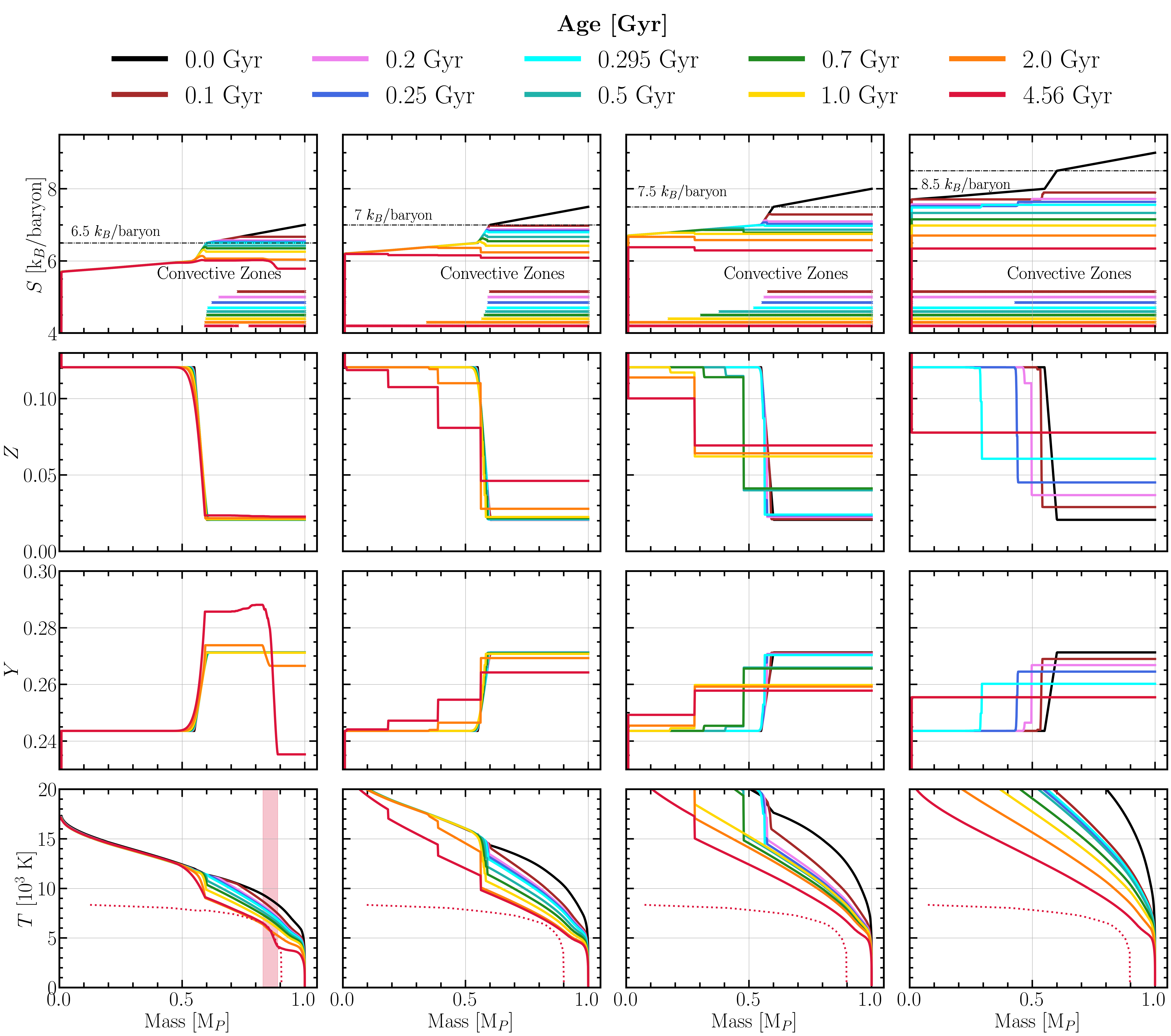}
\caption{Example evolution models of four mock models with varying initial entropy profiles, increasing from low entropy (left column) to high entropy (right column). These models were initialized with a steep $Z$ profile gradient surrounded by flat $Z$ regions, but are provided with stabilizing $S$ profile gradients on either side. To observe the effect of varying the overall entropy profile, these models all begin with the same $Z$ profile, with a steep gradient fixed initially for all models at 0.55-0.6 \Mj. 
From left to right, this figure shows that higher internal entropies destroy the initial $Z$ profile gradients.
As a result of their lower initial entropies, those that survive experience colder temperatures in the convective region. 
The first row shows the inward-moving convective zones as each model cools. 
The second row shows the evolution of the $Z$ profile.
Even a modestly cold initial interior entropy of 7.5 \kbbar can convectively mix an initially stable region at this location (0.6 \Mj). 
Higher interior entropies mix the $Z$ gradients to varying degrees, with the last model on the left mixing completely in under 200 Myr. 
The third row depicts the evolution of the helium abundance ($Y$) profile, including helium rain, where we have modified the demixing temperatures of the SR18 miscibility curves by +550 K for example purposes only.
Due to conservation of mass and the setting of the hydrogen/helium mass ratio equal to the protosolar value, regions of high $Z$ correspond to regions of low $Y$, and vice-versa. 
The last row shows the temperature profiles with the helium rain region shown as the red shaded region.
The two coldest models shown in the left-most columns experience less convective mixing over evolutionary timescales, and their outside regions are therefore allowed to cool without constant resupply of heat from the interior.
This leads to even colder outer temperature profiles, which eventually intersect the miscibility curve employed here (seen in gray).
On the other hand, the two hottest models shown in the rightmost columns mix completely and remain hotter throughout their interior leading to higher temperature profiles at later ages.
}

\label{fig:fig4}
\end{figure*}

\begin{figure}[ht!]
\centering
\includegraphics[width=0.47\textwidth]{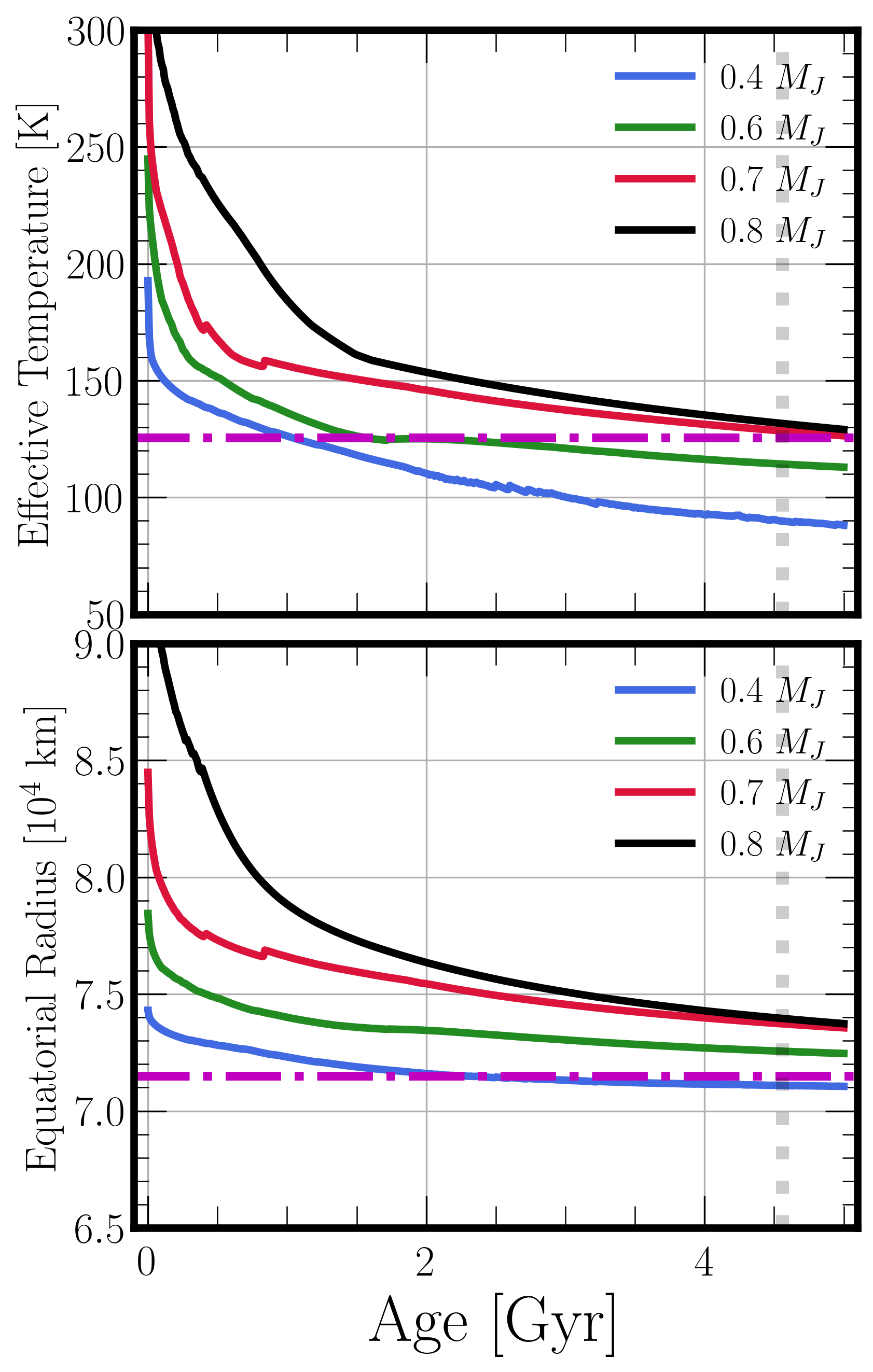}
\caption{Evolution of the effective temperature (top panel) and equatorial radius (bottom panel) of the four mock models presented in Figure~\ref{fig:fig4}. The four models are distinguished by the entropy at the boundary of the fuzzy core corresponding to the columns of Figure~\ref{fig:fig4}. The models initialized with higher entropies result in hotter effective temperatures and larger radii since their stable regions are convectively mixed at earlier ages. For example, the model with interior characteristic entropy of 8.5 \kbbar that the black line represents mixes entirely in under 200 Myrs. The red line representing the 7.5 \kbbar model mixes almost entirely by 1 Gyr, after which the effective temperature recovers from its steep initial dive. The discontinuity jumps visible in the red line are due to rapid mixing events of the $Z$ profile. The dotted gray line marks 4.56 Gyr. At later times, the effective temperature of the coldest model (blue line; 6.5 \kbbar) has cooled to under 100 K with a smaller radius than the rest of the models. This model was initialized with colder interior temperatures and thus did not convectively mix. Its convective region therefore evolved without heat resupply from the interior leading to low temperatures on the outside. The violet horizontal line in the upper panel marks the observed effective temperature of Jupiter of 125.57 K \citep{Li2012} and the violet line in the bottom panel the equatorial radius of Jupiter \citep[71,492 km;][]{Seidelmann2007}.}
\label{fig:fig5}
\end{figure}

\begin{figure*}[ht!]
\centering
\includegraphics[width=\textwidth]{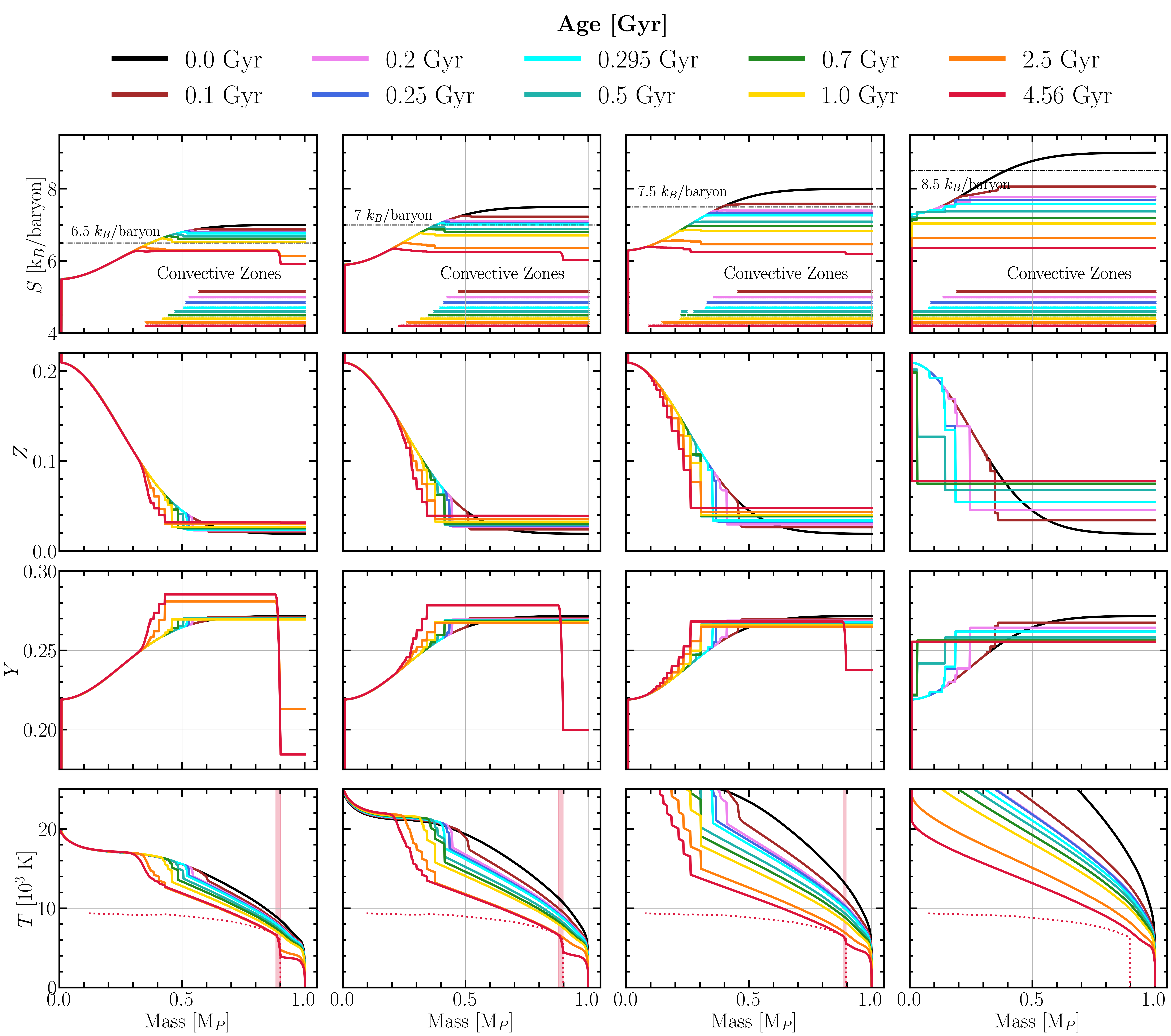}
\caption{Example evolutionary models with gradually extending cores. This figure follows the exercise conducted in Figure~\ref{fig:fig4}, starting at a median low interior entropy of $6.5$ \kbbar (left) to high median interior entropy of $8$ \kbbar (right). To show helium rain in these models and only for example purposes, we have used here an unmodified LHR0911 miscibility curve. As in all prior figures, the models have a total heavy envelope mass of 27 $\Mearth$ with a 2.5 $\Mearth$ compact core. The first row depicts the inward motion of the convective zones as the planet cools. The second row shows the evolution of the $Z$ profile and, importantly, showcases that higher interior entropies lead to rapid convective mixing of the composition gradient. Cooler initial entropy profiles let their outer regions cool at a faster rate without access to heat from the interior, leading to encounters with the LHR0911 miscibility curves at later ages. The third row shows the evolution of the total helium abundance, $Y$, and its depletion due to helium rain. Given mass conservation, regions of higher heavy element abundance correspond to regions of lower helium abundance. 
The last row shows the evolution of the temperature profile and the helium rain region is shaded in red. 
Compared to the models presented in Figure~\ref{fig:fig4}, these initial conditions keep the convective zones from moving to the deep interior regions, leading to stable regions at Gyr timescales, with higher metal abundances in the exterior without mixing completely.} 
\label{fig:fig6}
\end{figure*}

\begin{figure}[ht!]
\centering
\includegraphics[width=0.47\textwidth]{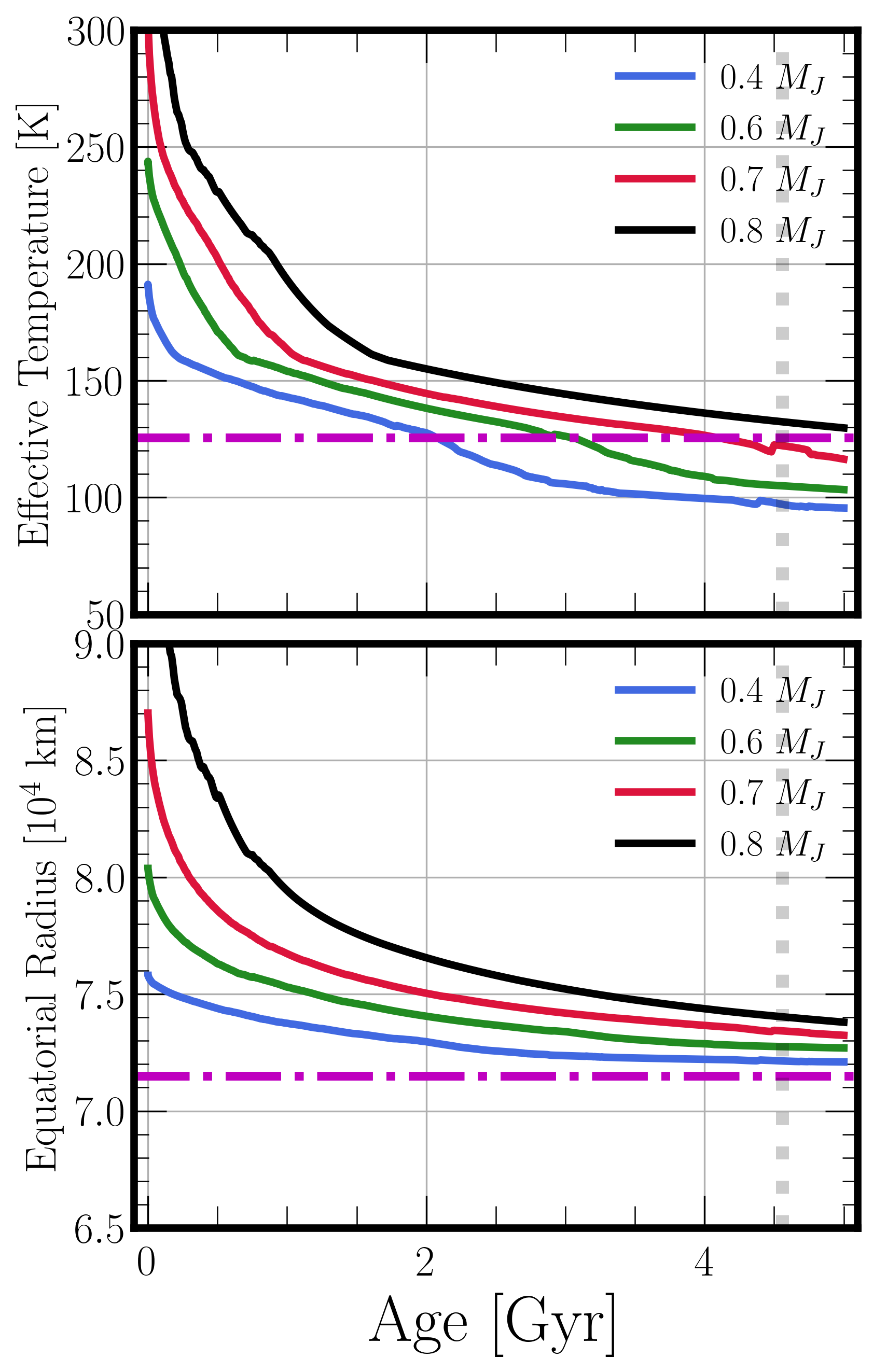}
\caption{Evolution of the effective temperature (top panel) and radius (bottom panel) of the four models presented in Figure~\ref{fig:fig6}, denoted by the corresponding median internal entropies corresponding to the columns of Figure~\ref{fig:fig6}. The dotted gray line marks 4.56 Gyrs. Unlike the models in Figure~\ref{fig:fig4}, the initial distributions of $S$ and $Z$ allow for a more gradual evolution of the effective temperature even for the 7.5 \kbbar model, represented by the red curve. This model retained a stable region out to 0.35 \Mj, and came closer to Jupiter's effective temperature.  The magenta line in the top panel marks the observed effective temperature of Jupiter of 125.57 K \citep{Li2012} and the equatorial radius of Jupiter \citep[71,492 km;][]{Seidelmann2007} in the bottom panel.}
\label{fig:fig7}
\end{figure}

\begin{figure*}[ht!]
\centering
\includegraphics[width=0.99\textwidth]{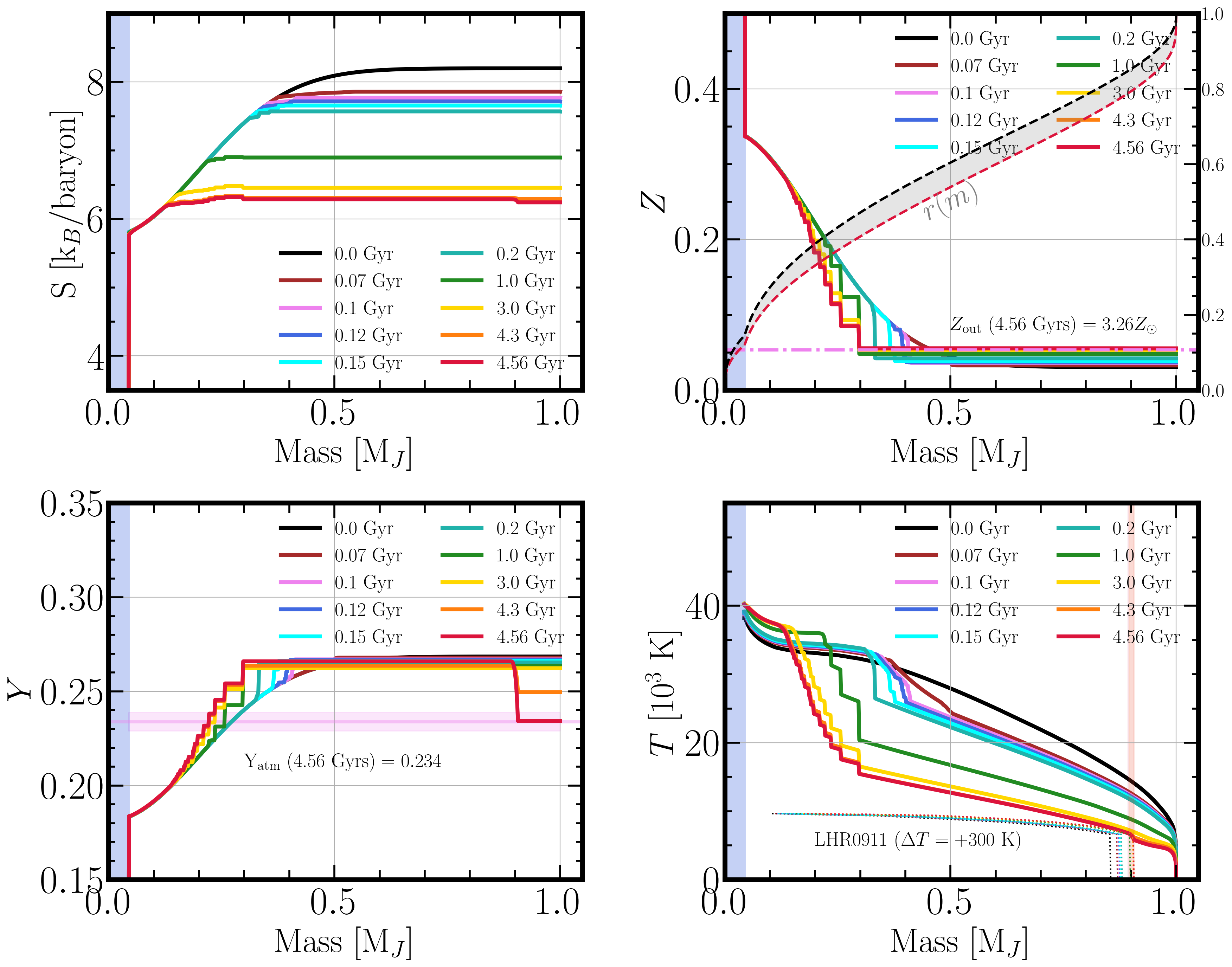}
\caption{Evolutionary model to the present-day that best matches the effective temperature, radius, atmospheric helium abundance, and outer envelope metallicity within observational uncertainties. The initial mean entropy per baryon is 7.5. The top left shows the evolution of the entropy profile, the top right the evolution of the composition ($Z$) profile, the bottom left the evolution of the helium ($Y$) profile, and the bottom right the evolution of the temperature profile. This particular model contains a total of 44 $\Mearth$ of heavy elements with a 14 $\Mearth$ compact core ($Z = 1$), marked by the shaded blue regions. As shown in the upper right panel, the initial $Z$ profile becomes convectively mixed from the outside toward the inside, and this change is most pronounced at early times. The dark-shaded region represents the normalized radius as a function of normalized mass, whose coordinates are on the second Y-axis and whose boundaries are color-coded according to the initial and final age. The resulting extended core ($dZ/dm < 0$) extends out to 30\% of the planet's mass, corresponding to 42\% of the radius at 4.56 Gyrs. The outer metallicity begins at 1.5$\Zsol$ and rises to 3.26 $\Zsol$. Regions of higher metallicity correspond to regions of lower helium mass fraction. Helium rain begins at 4 Gyrs in this model, for which we use the LHR0911 miscibility curves with a temperature shift of +300 K to deplete the helium in the outer layers to 0.234. The latter is consistent with the \textit{Galileo} entry probe measurements \citep{vonzahn1998}. The demixing temperatures encountered by the temperature profile are shown in the last panel. The dotted lines in the last panel show the adjustment of the LHR0911 miscibility curves to their corresponding temperature and pressure profiles. The long-shaded region corresponds to the helium rain region.}
\label{fig:fig8}
\end{figure*}

\begin{figure}[ht!]
\centering
\includegraphics[width=0.47\textwidth]{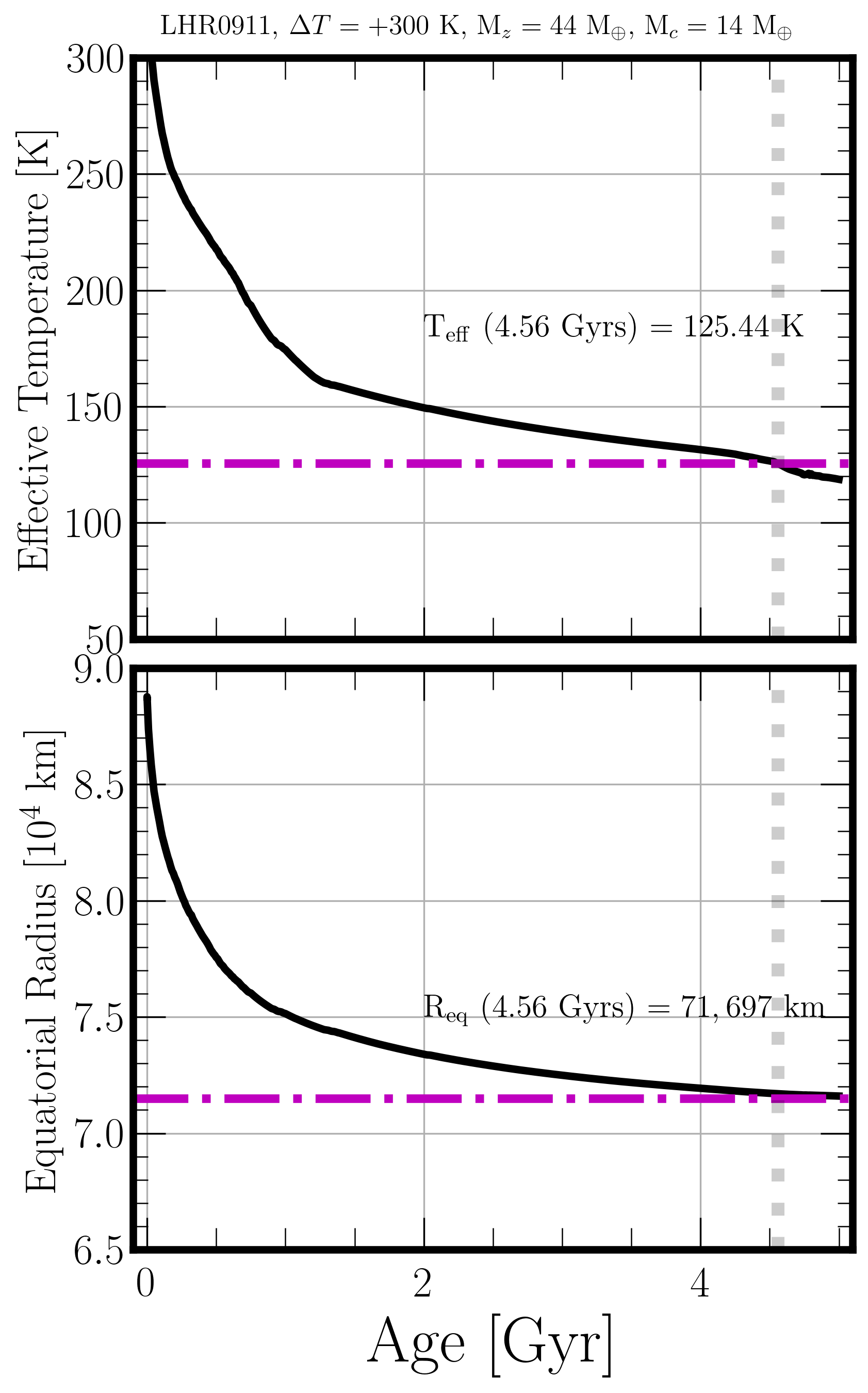}
\caption{Evolution of the effective temperature (top panel) and equatorial radius (bottom panel) of the preferred evolutionary model of Jupiter in Figure~\ref{fig:fig8}. In the top panel, the model reaches the observed effective temperature of Jupiter \citep[$125.57 \pm 0.07$ K;][]{Li2012} to within 0.06 K (0.05\%) of the observational uncertainty. The effective temperature of Jupiter is shown as the horizontal magenta line in the top panel. This model comes to within 0.2\% of the equatorial radius of Jupiter \citep[71,492 km;][]{Seidelmann2007} at 4.56 Gyrs. The values of Jupiter compared to those of this model are provided in Table~\ref{tbl2}. The equatorial radius is shown as the dash-dot magenta line in the bottom panel.}
\label{fig:fig9}
\end{figure}

\begin{figure}[ht!]
\centering
\includegraphics[width=0.47\textwidth]{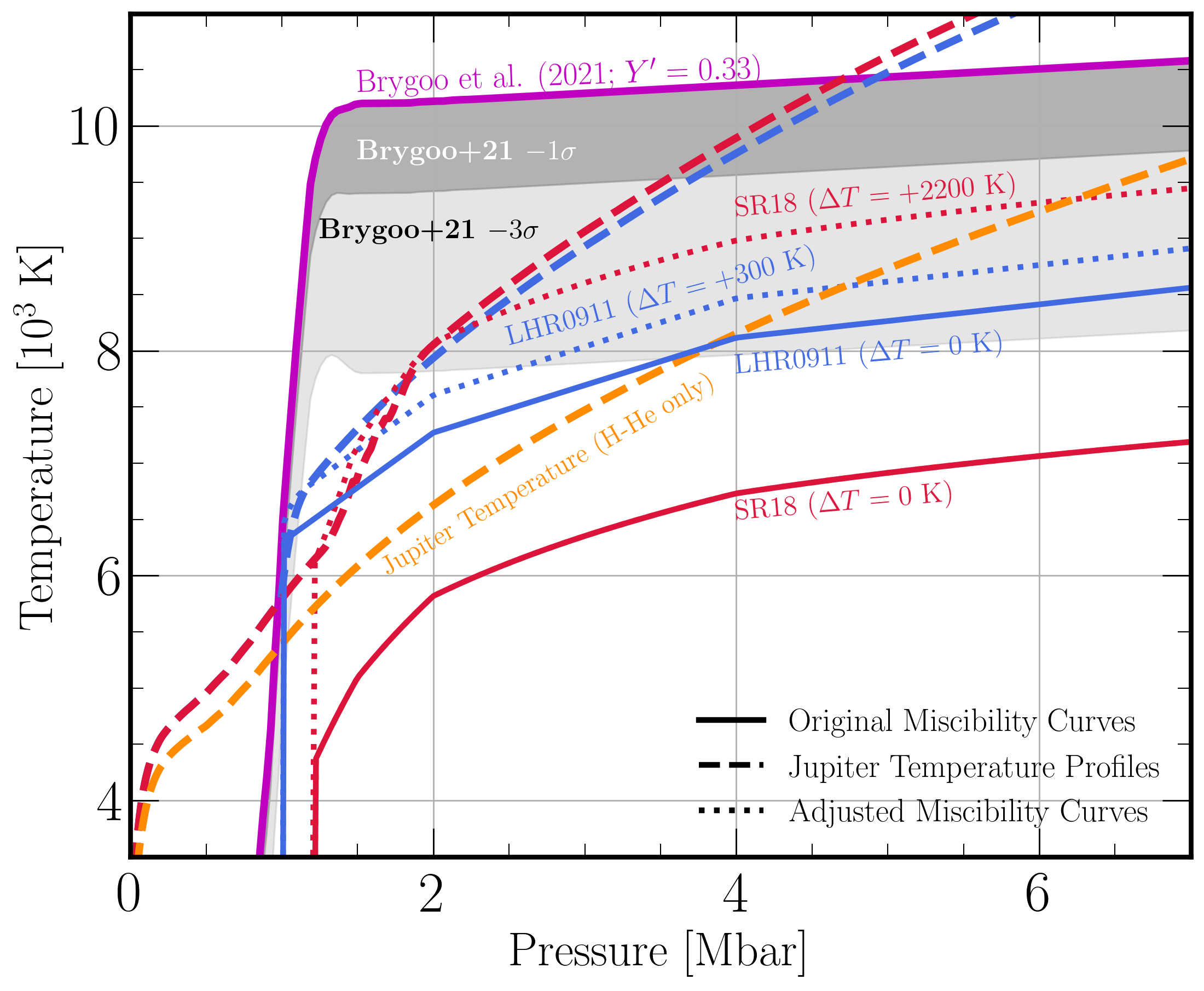}
\caption{Comparison of the original LHR0911 and SR18 miscibility curves (solid lines) with two Jupiter temperature profiles at 4.56 Gyrs, including the effects of the inclusion of heavy elements in the EOS. The low-pressure demixing temperatures of the SR18 miscibility curves have been limited to 1.2 Mbar and those of LHR0911 to above 1 Mbar. The dotted blue curve depicts our favored model using the LHR0911 miscibility curve, the dotted red curve is a similar model to the SR18 miscibility curve, and the dashed yellow model is a fiducial H-He homogeneous Jupiter profile model. At this age, the outer envelope has a $Z$ of  $\sim$3.3 $\Zsol$, making the temperatures significantly higher compared to those of a 1$\Zsol$ model and more so when compared with the pure H-He model (dashed yellow). These higher temperatures at 4.56 Gyrs motivate the need for higher miscibility temperature modifications compared to the recent H-He models of \citet[][+539 K, SR18]{Mankovich2020} and \citet[][-1250 K, LHR0911; +350 K, SR18]{Howard2024}. The interpolated experimental result of \cite{Brygoo2021} is shown in magenta for comparison. The darker and lighter shaded region highlights the lower 1$\sigma$ and 3 $\sigma$ uncertainties of that experiment, respectively.}
\label{fig:fig10}
\end{figure}

Generally, the $Z$ profiles that retain stable regions are deeper in the planet ($\leq 0.4$ M$_J$) compared to those inferred by DC19, MH22, and MH24 and must begin with a characteristic interior entropy of $\sim$7.5 \kbbar, which is compatible with cold-start formation models \citep{Cumming2018, muller2020}.
Lower interior entropies lead to colder outer temperatures, and higher interior entropies lead to more convective mixing (see Figure~\ref{fig:fig4}). In addition, metallicity profile gradients surrounded by flat profiles located at $\gtrsim 0.4\ \Mjup$ will either convectively mix or lead to effective temperatures significantly colder than 125 K. 
Including a stabilizing region interior to the steep profile gradient leads to a more stable core but aggravates the fit with effective temperature. 
As a result, initial $Z$ distributions with steep stable profile gradients located in the outer regions (M $\geq$ 0.6 M$_J$) may not be the result of formation processes.

Obtaining $\sim$3$\Zsol$ metallicity in the outer envelope leads to higher temperatures in the outer convective region since (generally) higher metallicity temperature profiles are significantly hotter than pure H-He profiles. 
As a result, to deplete sufficient helium from the atmosphere, significantly higher H-He demixing temperatures are required. In this work, the LHR0911 and SR18 miscibility curves require higher temperatures by $\sim$300 K and $\sim$2200 K,  respectively. 

While the observed metallicity of Jupiter's atmosphere may be due to recent impact events \citep[see discussions in][]{Debras2019, Helled2022b}, it remains unlikely that positive $Z$ profile gradients will remain stable over even megayear timescales. 
As such, we have assumed here that the atmospheric metallicity of Jupiter is the same throughout the outer convective region.

We emphasize that the fits to the \textit{Juno} gravity data constrain only the mass density profile in the current Jupiter. 
The interior thermal distributions are not constrained at all and are therefore assumed by all researchers attempting to use the \textit{Juno} gravity data. 
This accounts for the wide range of inferred density distributions and forcefully suggests that the interior thermal profiles are very much unconstrained. 
This is where evolutionary models can play a useful role. 
Since they fundamentally depend upon the thermal interior and the associated energy transport model, they bring to bear other constraints, such as the atmospheric effective temperature at the current epoch. 
They also can connect the formation phase with the current phase, something Nature does and theorists should attempt. 
However, our current models shed little light on Jovian birth, other than to constrain the possible character of a fuzzy core at formation, if that measured extended core is primordial. 

We have left to future work attempts to fit in the context of evolutionary models the current \textit{Juno} lower-moment ($J_2$ and $J_4$) gravity data (Sur et al. 2024 b, \textit{in prep.}). 
This is in part due to the difficulty researchers have had fitting them in the context of Jupiter's measured surface metallicity \citep{Nettelmann2017, Nettelmann2021, Helled2022b}. 
Such static fits to \textit{Jupiter} (see Figure~\ref{fig:fig1}) seem often to require low metallicities, and sometimes negative metallicities \citep{Nettelmann2021} in the outer envelope.
\cite{Debras2019} has proposed a positive $Z$ gradient in the outer regions of Jupiter (red line in the bottom panel of Figure~\ref{fig:fig1}) 
while \citet{Howard2023c} find that a positive $Z$ gradient of such magnitude requires an excessive amount of late accretion inconsistent with collisional evolution models of the Solar System.
On the other hand, \cite{Muller2024} argue that a possible reduction in local opacity in the deep atmosphere could form a stable layer even in the presence of a destabilizing $Z$ gradient shortly after formation, allowing the upper atmosphere to be enriched by collisions, while the deep interior remains metal poor.

As proposed by \cite{Stevenson1982}, core erosion could be responsible for generating $Z$ gradients and fuzzy cores on evolutionary timescales, and core erosion will be a subject of future work. 
We have not in this paper incorporated the potential effects of semi-convection and semi-convective interfaces \citep{Moll2016, Moll2017, Fuentes2022, Tulekeyev2024} that have been shown to endure over evolutionary timescales in the presence of rotation \citep{Hindman2023, Fuentes2023, Fuentes2024}, though \citet{muller2020} found they had no significant effect without rotation.  
Clearly, the effects of both rotation and semi-convection require further scrutiny.

Uncertainties exist in the hydrogen-helium and heavy-element equations of state, the H-He demixing temperatures and pressures, and in other material properties at high pressures that render both evolutionary and static models provisional. In the context of the high $Z$s inferred in current models, the lack of physically sophisticated equations of state for hydrogen/helium/heavy-element mixtures for the range of compositional mixtures discussed in the literature is a particularly important shortcoming of all current models of Jupiter (and Saturn).

Despite all these uncertainties, evolutionary models are crucial for determining the long-term stability of these fuzzy cores in the presence of convective mixing, in informing formation models, and in clarify what evolutionary paths are feasible, given the various observational constraints.
Their continued development and refinement are key to the eventual understanding of this class of planet, whose importance extends beyond the Solar System giants to the giant exoplanets now being discovered in profusion and that await a definitive understanding of our local variants.

\clearpage

\begin{deluxetable*}{ccc}
\tablewidth{0pt}
\tablecaption{Jupiter Data vs. Derived Evolutionary Model}
\tablehead{
\colhead{Parameter} & \colhead{Value} & \colhead{Derived Model Values} 
}
\startdata
$\Teff$ [K] & $125.57 \pm 0.07$ & $125.44$\\
$\Yatm$ & $0.234 \pm 0.005$ & 0.234\\
$Z_{\rm out}$ [$Z_{\odot}$] & $\sim$3 & 3.26\\
Equatorial Radius [km] & $71,492$ & $71,697$\\
\enddata
\tablecomments{Jupiter observable data compared to the values of our preferred model from \S\ref{sec:results} at 4.56 Gyrs. This particular model has a total heavy element mass of $\Mz = 44\ \Mearth$ and a compact core mass of $\Mc = 14\  \Mearth$ and uses the LHR0911 miscibility curve with a demixing temperature modification of +300 K to reach the observed helium mass fraction of 0.234.}
\end{deluxetable*}
\label{tbl2}

\begin{deluxetable*}{cc}
\tablewidth{0pt}
\tablecaption{H-He Miscibility Temperature Shifts}
\tablehead{
\colhead{Authors} & \colhead{Misc. Curve ($\Delta T$)}
}
\startdata
This Work & LHR0911, SR18 ($+300$ K, $+2200$ K) \\
\cite{Howard2024} & LHR0911, SR18 ($-1250$ K, $+350$ K)  \\
\cite{Mankovich2020} & SR18 ($+539$ K) \\
\cite{Mankovich2016} & LHR0911 ($ \lesssim -100$ K) \\
\cite{Pustow2016} & LHR0911 ($+500$ K) \\
\cite{Nettelmann2015} & LHR0911 ($+250$ K) \\
\enddata
\tablecomments{List of H-He miscibility temperature shifts in the literature needed to match helium abundance observations of Jupiter and Saturn. The temperature shifts of this work as a result of evolving stably stratified $Z$ profiles are provided in the first row.
We note that the experimental results of \cite{Brygoo2021} show a demixing temperature of H-He that is $\sim$7000 K higher than the SR18 demixing temperatures and $\sim$3000 K higher than the LHR0911 demixing temperatures at megabar pressures. }
\end{deluxetable*}
\label{tbl3}

\begin{acknowledgments}
We thank the referees for their thorough reviews that very much sharpened numerous aspects of the manuscript. Funding for this research was provided by the Center for Matter at Atomic Pressures (CMAP), a National Science Foundation (NSF) Physics Frontier Center, under Award PHY-2020249. Any opinions, findings, conclusions, or recommendations expressed in this material are those of the author(s) and do not necessarily reflect those of the National Science Foundation. 
RTA is supported by the Ford Foundation Predoctoral Fellowship.
YS is supported by a Lyman Spitzer, Jr. Postdoctoral Fellowship at
Princeton University. We thank Dr. Matthew Coleman and Princeton Research Computing for their computational and technical support. We further thank Yi-Xian Chen, Dr. Jisheng Zhang, and Dr. Akash Gupta for useful discussions. 
\end{acknowledgments}

\clearpage
\appendix

\begin{figure}[ht!]
\centering
\includegraphics[width=\textwidth]{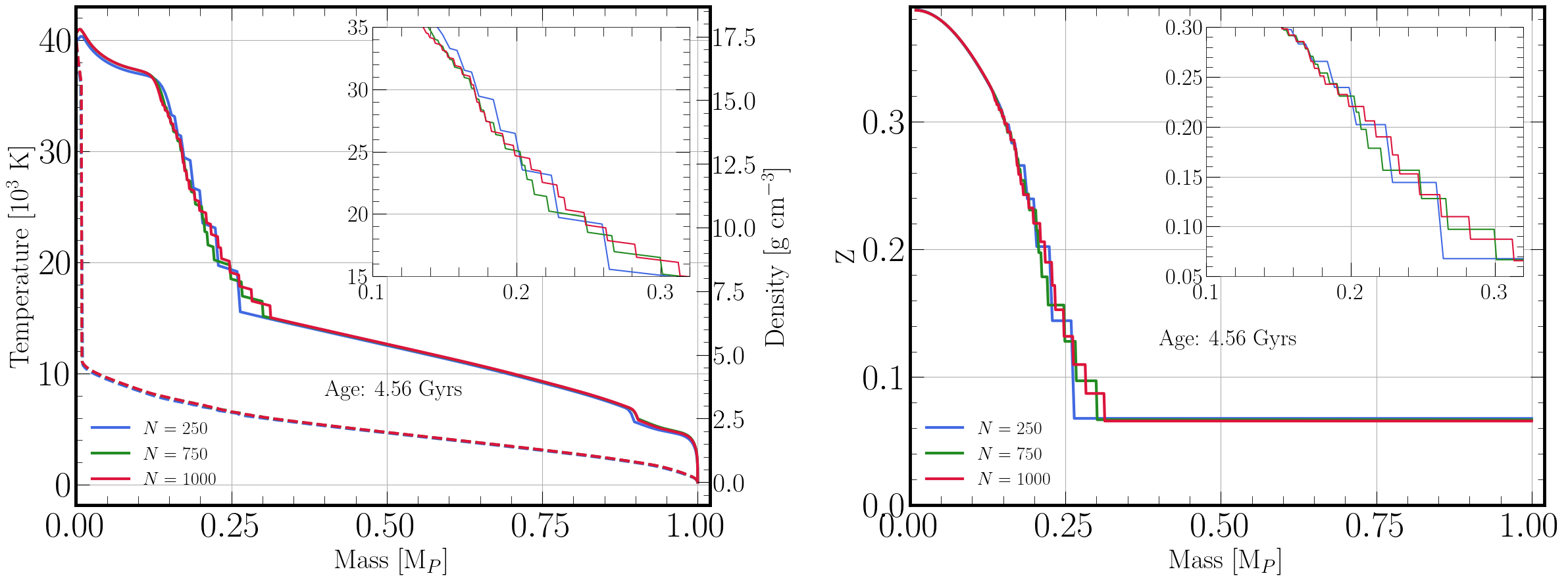}
\caption{Left: The temperature and density profiles of different models with different spatial resolutions ($N$) of 250, 750, and 1000 zones, all at 4.56 Gyrs. Right: Heavy metal distribution of the same models at the same age. The inset axes shows the size differences of the staircases. Overall, the number of mass zones used influences the size and number of staircases, but not their location. All models exhibit the staircases at roughly the same $Z$ abundances, and the outer region that experiences homogenization by convection is also of an extent that is independent of the number of zones.}
\label{fig:fig11}
\end{figure}

Below, Figure~\ref{fig:fig11} shows example Jupiter profile models at the present epoch with 250, 750, and 1000 mass zones. The size of the staircases is affected by the model spatial resolution. The location and shape of the stable region, however, is not significantly affected by the spatial resolution choice. The choice of time resolution is even less significant than the variations of spatial resolution shown here.

\clearpage

\bibliography{references_free}{}
\bibliographystyle{aasjournal}

\end{document}